# Gender-Based Violence in 140 Characters or Fewer:
# A #BigData Case Study of Twitter

Hemant Purohit[1,2], Tanvi Banerjee[1,2], Andrew Hampton[1,3], Valerie L. Shalin[1,3],
Nayanesh Bhandutia[4] & Amit P. Sheth[1,2]

[1]Ohio Center of Excellence in Knowledge-enabled Computing (Kno.e.sis), Dayton, OH, USA
[2]Department of Computer Science and Engineering, Wright State University, Dayton, OH, USA
[3]Department of Psychology, Wright State University, Dayton, OH, USA
[4]United Nations Population Fund Headquarters, NYC, NY, USA

**Corresponding Authors:**
Hemant Purohit, Amit Sheth, Valerie Shalin: {hemant, amit, valerie}@knoesis.org

## ABSTRACT

Public institutions are increasingly reliant on data from social media sites to measure public attitude and provide timely public engagement. Such reliance includes the exploration of public views on important social issues such as gender-based violence (GBV). In this study, we examine big (social) data consisting of nearly fourteen million tweets collected from Twitter over a period of ten months to analyze public opinion regarding GBV, highlighting the nature of tweeting practices by geographical location and gender. We demonstrate the utility of Computational Social Science to mine insight from the corpus while accounting for the influence of both transient events and sociocultural factors. We reveal public awareness regarding GBV tolerance and suggest opportunities for intervention and the measurement of intervention effectiveness assisting both governmental and non-governmental organizations in policy development.

**KEY FINDINGS & IMPLICATIONS**

*Social Media Content:*
- Substantial GBV related content exists in social media.
- Spikes in GBV content reflect the influence of transient events, particularly involving celebrities.
- Gender, language, technology penetration, and education influence participation with implications for the interpretation of quantitative measures.
- GBV content includes humor and metaphor (e.g. in sports) that reflect both attitude and behavior.
- Content highlights the role of government, law enforcement and business in the tolerance of GBV.

*Relevance to GBV Policy:*
- Social Media provides an alternative for measuring GBV attitude and behavior that is cheaper, faster, and broader than conventional survey-based methods.
- Regional socio-cultural context influences both the measurement and the interpretation of data.
- Computational methods for context-sensitive monitoring, modeling and interpreting GBV social media content are highly feasible.
- Sociocultural variability necessitates regional metrics and anti-GBV campaigns.
- Policy makers require tools to make social media content accessible in near-real time to monitor the effectiveness of anti-GBV campaigns.

**[1.] INTRODUCTION**

Gender-based violence (GBV), primarily against women, is a pervasive, global phenomenon affecting both developed and developing countries. Over 35% of the world's female population has experienced gender-based violence at some point in their lives (World Health Organization, 2013). According to the United Nations Population Fund (UNFPA)[1], "GBV is a serious public health concern that also impedes the crucial role of women and girls in development." However, anti-GBV sentiment is not universal, apparent in sexist chants during professional sport[2] and public response to convicted offenders[3]. While a United Nations campaign acknowledges GBV as a societal problem (United Nations, 2009), GBV prevalence is difficult to assess. The European Union's council report highlighted a persistent lack of comparable data across regions and over time (European Union, 2010), hampering both assessment and mitigation. Both the UNFPA[4] and the European Union Agency for Fundamental Rights[5] seek better data sourcing and policy design (European Union Agency for Fundamental Rights, 2012).

Here we examine the utility of Computational Social Science to address the problem of monitoring public views and mitigating GBV on a global scale (see Table 1). Mining large-scale online data from mobile technology and social media such as Twitter promises to complement traditional methods and provide greater insight with finer detail. Computational Social Science (Lazer et al., 2009) has leveraged such data to inform programs in a variety of domains, for example disaster response coordination (Purohit et al., 2014; Purohit et al., 2013; Vieweg et al., 2010), health (Culotta, 2014; De Choudhury et al., 2013), and drug abuse (Cameron et al., 2013).

The existing studies of Twitter in the GBV domain provide an exploratory analysis of misogynistic language usage in developed countries, the UK and US (Bartlett et al., 2014; Fulper et al., 2014). While some of our conclusions are similar to Bartlett et al. 2014 who focused on the UK, their specification of methods (regarding location and gender identification, and diagnostic features of language) is limited. Fulper et al. 2014 document a relationship between misogynistic language and rape statistics in the US, but shed little light on the content of such language. We complement these exploratory studies by collecting and analyzing a broader topical dataset (e.g., sexual violence and harmful practices) with respect to language usage beyond just misogyny. Consistent with the UNFPA, our interest lies in the development of GBV metrics for global monitoring. This requires a global dataset for developing countries beyond the UK and US, especially countries that have prevalent GBV issues, in order to incorporate regionally appropriate sociocultural factors. Below, we provide analyses of tweet volume, location, timing, and source author's gender along with several content analyses to help monitor GBV. Both inform the design and assess the effectiveness of anti-GBV campaigns and policy. **In so doing, we**

---

[1] UNFPA agency's description about GBV: http://www.unfpa.org/gender/violence.htm
[2] http://www.bbc.com/news/blogs-trending-31628729
[3] http://www.telegraph.co.uk/news/worldnews/asia/india/11443462/Delhi-bus-rapist-blames-his-victim-in-prison-interview.html
[4] UNFPA agency: http://www.unfpa.org/public/
[5] EU FRA agency: http://fra.europa.eu/en/project/2012/fra-survey-gender-based-violence-against-women

**identify the computational feasibility of addressing many of the challenges of traditional GBV monitoring.**

We review the challenges of traditional monitoring to provide a benchmark for assessing the success of our approach. Gathering statistics on GBV episodes is time consuming, collected under non-standardized protocols, and published in highly aggregated form. For example, the non-partner sexual violence prevalence data published by the World Health Organization[6] is dated 2010 and reported by 21 aggregated geographic regions such as West Africa and South Asia. While the United Nations Office on Drugs and Crime data[7] (consisting of trafficking and sexual violence data) are less aggregated, and somewhat more recent with some data available as recently as 2012 and separated by country, the data are sparse and incomplete. For instance, sexual violence data are available for the Philippines and Nigeria for 2012, but the most recent data for India are from 2010 and no data are presented for South Africa. The lag itself prevents monitoring change, either to detect unexpected increases in GBV, change in attitude possibly due to recent events and mitigation efforts. Moreover, the conventionally available statistics reported above reflect legal definitions, making direct comparisons among countries impossible due to differences in the definition and recording of offenses.

*Table 1*. Tweet samples from our analysis (see Section 3) with implications to inform and design targeted GBV intervention campaigns. M1-M3 illustrate existing automatic analysis capability, while M4-M8 result from partially automated analyses presenting the case for expanding computational methods.

| Message | Implications |
|---|---|
| *M1:* RT @USER1: 1 in 3 women are raped/abused in their lifetime. RT if you rise to stop the violence. #1billionrising http://t.co/lXEEmQoLbO | *Volumetric analysis* of social media can help measure population engagement and effective penetration of designed campaigns in the community |
| *M2*: #StopRape Rape Crisis says many survivors of sexual abuse and assault still don't feel confident in the criminal justice system. CS | *Location analysis* permits the identification of y message origin (e.g., South Africa in M2) and routing to appropriate agencies |
| *M3:* @USER2 Takes a MOMENT 2 Sign & ask Others 2,so DV & Rape Laws become Equal. TOGETHER we can Change History: http://t.co/ylWtl5PgCI | *Gender detection analysis* of message authors (e.g., female in M3) supports adjustment of apparent population GBV attitudes by gender and suggests the content of anti-GBV policy and campaigns |
| *M4:* RT @USER3: Rape prevention nail polish sounds like a great idea but Iím not sure how you're going to get men to wear it | *Content analysis* sensitive to sociocultural considerations permits the assessment of subtle measures such as the role of humor, sarcasm and despair (e.g., in M4 from the Philippines) |
| *M5:* RT @USER4: 15 yard penalty for "unnecessary rape" http://t.co/yhzxtYzGP0 | *Metaphor analysis* indicates the acceptance of GBV, echoed for example in sports, and suggesting opportunities for specific anti-GBV campaigns |
| *M6:* Valentine's day is really helping me sell | *Entity recognition* based on knowledge-bases of GBV |

---

[6] Data available here: http://apps.who.int/gho/data/view.main.NPSVGBDREGION
[7] Data available here: http://www.unodc.org/unodc/en/data-and-analysis/statistics/crime.html

| these <u>date rape drugs</u> | entities can help identify precipitating events (e.g., Valentine's day in M6), apriori to design preventive campaigns |
|---|---|
| **M7:** @USER5 <u>WB Govt</u> must have <u>ordered</u> Police to <u>protect the family of Rape-Victim</u>. It is <u>shameful for Mamata Banerjee</u> O GOD GIVE WISDOM TO ALL | *Organization detection*, including government, law enforcement and commercial entities in relation to GBV can inform policy, e.g., M7 informs potential lack of police protection for a victim's family in West Bengal (WB) |
| **M8:** RT @USER6: It is not my job to coddle and <u>"educate" young Black men</u> when it comes to <u>violence against women</u>. Y'all wanna "teach"? | *Modeling of stereotypical association* can inform design of targeted campaigns, e.g., M8 author is stereotyping GBV violence with black men |

*Note.* We anonymized user mentions as per the IRB guidelines.

Apart from the logistical problems of gathering GBV data, the data have validity limitations. Reliance on formal reports to law enforcement risks under-reporting by victims and witnesses who may believe that domestic violence is a private matter (Nilan et al., 2014). Aggregation/generalization across localities with different socio-economic properties masks important trends (Yoshihama, Blazevski, & Bybee, 2014) because sociocultural context including politics, history, religion, and economy strongly influence attitudes (Nayak et al., 2003). Researchers suggest the need for different prediction and mitigation models for different sociocultural contexts (Corroon et al., 2014; Martinez & Khalil, 2012).

Social science survey research methods complement the law enforcement data.. However, social science survey data also has several limitations. Surveys reflect sampling biases and various confounds (Pierotti, 2013; Kalichman et al., 2005; Nilan et al., 2014). We highlight four limitations of survey data of particular relevance to Computational Social Science methods. First, survey items tend to address attitudes but not behavior, and therefore bear unclear relationship to the rate of GBV episodes (Bhanot & Senn, 2007). In contrast, inasmuch as verbal abuse constitutes a form of violence, social media posts can provide actual instances of this behavior. Second, the methods fail to account for transient global events, such as political or celebrity activity that can influence views and responses, hindering comparison over time. Third, the items themselves presume an established theory and standard measures of GBV, limiting the opportunity to discover latent patterns that reflect attitude and behavior. For example, metaphor is a powerful reflection of public opinion (Lakoff, 2008), but to our knowledge has not been explored in survey measures. Fourth, survey methods constitute a highly labor intensive data collection method, affording small samples while imposing cost and lag in data availability in a dynamic world (Pierotti, 2013). We will return to these issues when we evaluate our own approach. Social media promises a faster, cheaper, and face-valid means to engage the public, providing unprecedented large-scale access to public views and behavior (see Table 1). They provide an ability to monitor attitudes in near real-time and to support timely measurement and mitigation efforts. However, while use of social media provides speed, participation, and cost advantages, the absence of controlled sampling necessitates adjustments for demographics including literacy in internet infrastructure. Furthermore, we require region specific models to accommodate sociocultural differences in the definition and expression of GBV issues.

Ultimately, all three resources (formal reports, surveys, and social media) require integration in order to assist policy design, prioritize attention for interventions, and design region-specific programs to curb GBV. A logical first step is to demonstrate the potential of social media for GBV monitoring and the design of mitigation and policy.

**Study Design Overview**

Based on the suggestions of our UNFPA collaborators to identify a GBV-related corpus, we selected three major themes that encompass gender violence concerns: physical violence, sexual violence, and harmful sociocultural practices. Corresponding to these three themes, we created a seed set of keywords for data crawling from the Twitter Streaming API. We also selected four countries with suspected elevations in GBV suggested by UNFPA experts: India, Nigeria, the Philippines, and South Africa, in addition to the U.S., which had been subjected to preliminary analysis in Fulper 2014. Below, we examine tweeting practices by geography, time, gender, and events.

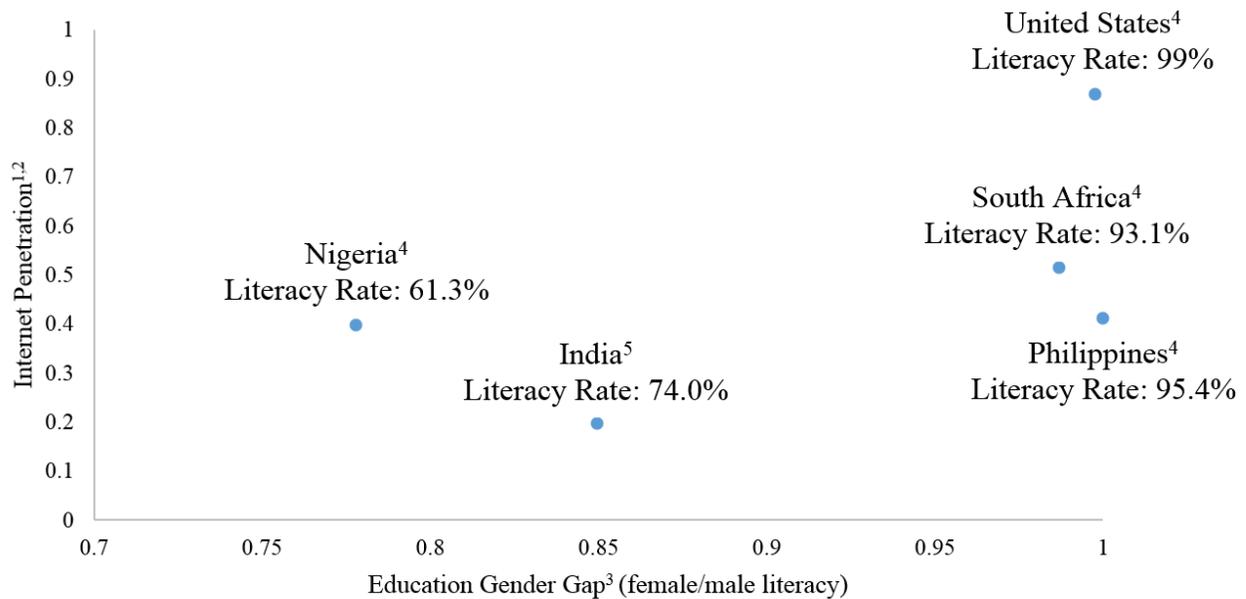

[1]http://www.internetworldstats.com/stats
[2]http://www.digitalstrategyconsulting.com/india/
[3]http://www3.weforum.org/docs/GGGR14/GGGR_CompleteReport_2014.pdf
[4]The World Factbook
[5]Registrar General and Census Commissioner of India

*Figure 1.* Summary of the education gap between genders, the penetration of Internet, and overall literacy rates in the diverse set of chosen countries.

The five countries present different contexts both for understanding social media data pertaining to GBV and for mitigation efforts. Figure 1 summarizes their variability across some key contextual dimensions: the education gap between genders, the penetration of the Internet, and overall literacy rates. The figure illustrates the clustering of Nigeria and India for lower literacy rates, a greater education gap, and lower Internet penetration. South Africa and the Philippines cluster with the U.S. regarding overall literacy and the reduced education gap, but reflect a diverging range of Internet penetration.

The graphs suggest the risk of sampling bias that affects data interpretation: an illiterate female citizen with no access to the Internet (likely in rural areas with unique GBV issues) may not be providing social media data, biasing the aggregated measures of attitude. Other differences not presented may also be relevant. For example, India is among the top 20 of over 140 countries regarding female political empowerment while Nigeria is below average on this dimension. Additional influences on the use of social media not depicted in the figure include cultural influences on free speech. For example, Nigerians may avoid public conversation about the Boko Haram atrocities[8] due to fear of revenge.

In the next section, we employ quantitative and qualitative analyses to examine Twitter content related to GBV. Twitter supports the distribution of short messages called tweets that are a maximum 140 characters in length. The character limit influences message style and constrains communication practices. Therefore, tweets often contain URL links to web pages or blogs, sometimes relying on shortened URL versions from external services (e.g., http://bit.ly/1C4HnMN). A hashtag convention (e.g., #RapeJoke, #ChildMarriage) supports the identification of searchable user-defined topics. Other Twitter engagement features include retweet (or 'RT', a forwarding of someone's tweet). The electronic device used to post a tweet may provide accessible, precise location indicators in some cases. Alternatively, accessible user profiles provide more general indicators of location and sometimes gender indicators such as author name.

A corpus of social data collected over a period of ten months included nearly fourteen million GBV tweets. In this corpus, we examine volume, location, trends over time, gender participation, and content such as metaphor and humor. Our analyses present both challenge and opportunity to study the phenomenon of gender-based violence. Challenges concern the need for computational methods to discern public perception and attitude from complex contextualized behavior. Opportunity lies in gaining fine-grained, region-specific insights concerning the prevailing GBV attitudes and related policies along with potential approaches to mitigation.

## [2.] METHOD

Below describe our data collection, followed by a description of the analysis approach.

**Data Collection**

Based on domain expert guidance we collected data from the Twitter Streaming API (Twitter Developers, 2014), using its 'filter/track' method for the given set of keywords. We used the keyword-based crawlers (data collectors) of our Twitris platform (Nagarajan et al. 2009, Sheth et al., 2014), where the Twitter Streaming API provides the relevant tweets for a given set of keywords. The crawling method contained processing for: calls to the filter/track method of the API for data collection, extraction of relevant metadata (e.g., tweet text, timestamp) and database storage. Consultation with UNFPA domain experts informed the required keyword lexicon (containing single or multi-word phrases) for input to the data crawlers. These experts assisted in the definitions and associated terminology corresponding to the three themes of interest for GBV study: physical violence, sexual violence, and harmful practices (see Table 2 for the keywords selected).

---

[8] For an overview, see the following: http://en.wikipedia.org/wiki/Boko_Haram

*Table 2*. Seed set of key phrases for GBV related data crawling.

| *Physical violence* | woman dragged, women dragged, girl dragged, female dragged, woman kicked, women kicked, girl kicked, female kicked, woman beat up, women beat up, girl beat up, female beat up, woman beaten, women beaten, girl beaten, female beaten, woman burn, women burn, girl burn, female burn, woman acid attack, women acid attack, girl acid attack, female acid attack, woman violence, women violence, girl violence, female violence |
|---|---|
| *Sexual violence* | sexual assault, sexual violence, rape, woman harass, women harass, girl harass, female harass, woman attacked, women attacked, girl attacked, female attacked, boyfriend assault, boy-friend assault, stalking woman, stalking women, stalking girl, stalking female, groping woman, groping women, groping girl, groping female |
| *Harmful Practices* | child marriage, children marriage, underage marriage, forced marriage, sex trafficking, woman trafficking, women trafficking, girl trafficking, female trafficking, child trafficking, children trafficking |

For each single keyword K, Twitter provided messages containing any form of the keyword #K, k, K. For the multi-word phrase K, the service provided messages that contained all of the terms in K, regardless of order. Each message is returned with associated metadata, containing various tweet-related and author profile-related characteristics, such as tweet origin location (latitude and longitude), time of posting, author profile description and location, number of author followers (users that subscribed to the author's updates), and followees (users to whose updates the author subscribed).

Location metadata was crucial to geographical analysis. We first checked if tweet origin latitude-longitude were available from the device used to send the tweet. Otherwise we resolved the author profile location (e.g., city, state or country name), if available, using calls to the Google Maps API service. Author profile location supplements the sparse (1.9%) device origin location metadata, apparent in the pilot analyses. Author profile location provides an explicit indication of nationality of particular interest to the analysis of region-specific GBV behavior and attitude. We used a bounding box of latitude-longitude for a country of interest to identify a country-specific tweet dataset. Using Genderize API (http://genderize.io), we collected genders for tweet authors. We first fetched the real names of the twitter authors using metadata for their Twitter handles. We extracted first names to detect author genders via calls to the Genderize API with first names as parameters.

**Analysis Approach**

With UNFPA guidance, we analyzed the data corpus of 13.9 million tweets over the ten months in non-uniform time slices, starting with a smaller pilot phase and adding two additional extended data collection periods due to encouraging results from the pilot data.

- SLICE 1: Jan 1 2014 - Feb 15 2014: 1.5 months; **Phase-1** (pilot phase)
- SLICE 2: Feb 15 2014 - June 31 2014: 4.5 months; **Phase-2**
- SLICE 3: July 1 2014 - Oct 31 2014: 4 months; **Phase-3**

To study the diverse set of data from the three phases, we employed mixed methods (Creswell, 2013) to reveal both patterns across the corpus as well as the content of specific contributions. The focus of *quantitative* analysis is to provide data-driven insights of activity patterns in the social media community by examining large-scale distributions of GBV content by geography, time, and gender. However, the recovery of the meaning of a pattern requires more fine-grained analysis than mere statistical distribution. Therefore, the focus of our *qualitative* analysis is to reveal attitudes and behaviors across different countries and between genders. In both cases, our interpretation relies on context, regarding current events and sociocultural considerations, ultimately supporting the need for context-sensitive computation for monitoring GBV content in social media.

## [3.] RESULTS
**Quantitative Analysis**

We discuss four types of quantitative analyses in this section with respect to volume, theme, information sharing, and gender.

**Volume analysis.** To begin our study, we sampled an initial slice of data for 1.5 months from all over the world, which contained 2.3 million tweets worldwide related to GBV. For brevity, we skip the broad descriptive statistics regarding this sample and instead summarize some key observations. Regarding our interest in location-specific GBV data, we note relatively few tweets with device-based location information (1.91%). However, author profiles provide location related information as well. Overall, more than half of the data (54.74%) for 1.3 million users had location information. The demonstrated feasibility of data collection motivated our more extensive data collection effort, spanning an additional 8.5 months.

More than 11% of 13.9 million tweets belonging to the five countries were examined here for volume analysis by country. Table 3 provides the composition of the full data set by country. We note more than five times the traffic in the U.S. and India relative to the other countries. Moreover, the observed frequency ranking differs from the population demographic[9] information for these countries (in descending order: India, the U.S., Nigeria, the Philippines, South Africa). We suspect that Internet penetration is influencing data collection. Due to these scale differences, a simple frequency graph for all of the raw data over time would mask variability in the Philippines, Nigeria, and South Africa with smaller populations and variable Internet penetrability. Figure 2, therefore, scales the raw data by two factors: population and Internet penetration. This allows us to consider the relative prevalence of GBV topics between countries and over time. Below we discuss some of the emergent patterns by country. We highlight the need to interpret these patterns with respect to a broad and complex knowledge base that includes current events, precisely the sorts of analyses that are best accomplished with computational tools.

---
[9] http://en.wikipedia.org/wiki/List_of_countries_and_dependencies_by_population

*Table 3.* Volume of GBV related tweets by country.

| Country of Origin | Number of Tweets |
|---|---|
| United States | 698077 |
| India | 549033 |
| Philippines | 105550 |
| Nigeria | 134671 |
| South Africa | 104430 |

*Note.* This table represents 1,591,761 (11%) of the total 13,942,592 GBV tweets collected globally.

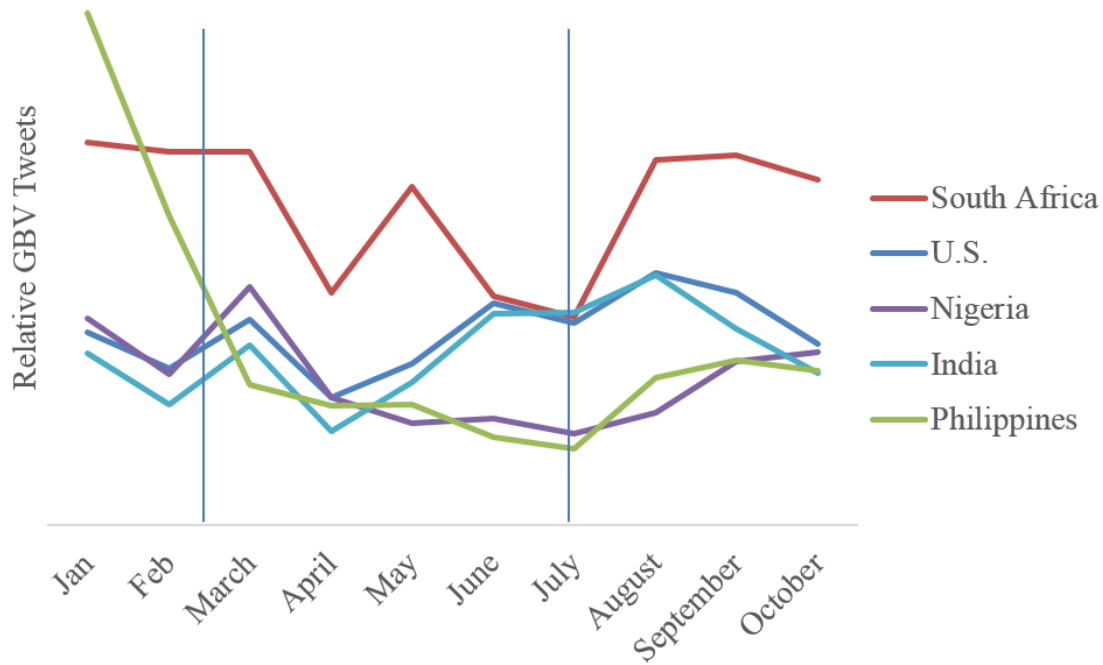

*Figure 2.* Tweet volume across countries from 1 January to 31 October 2014 related to GBV, and controlling for the population and Internet penetration of each country. Vertical lines represent time slices sampled. [Population expressed as a percentage of the smallest candidate (Nigeria). Nigeria = 1, India ~ 25. The adjusted metric, then, is expressed as original tweet numbers divided by adjusted population, the quotient is then divided by Internet penetration (the percentage of the population of each country that is online).]

Figure 2 appears to suggest that South Africa generates the most GBV related tweets overall, with the U.S. and India lagging behind roughly the same amount, and with Nigeria and the Philippines toward the bottom. However, we note that this cross-country pattern is sensitive to the form of adjustments made for population and Internet penetration.

Trends over time within a country, however, are not sensitive to such adjustments and are equally, if not more important. We make this point to illustrate the need to consider sociocultural influences on social media traffic. Several current events may explain the apparent peaks over time, revealed during a parallel Google Trends search by country. In the Philippines, the ongoing saga of Vhong Navarro which dominated much of the discussion[10]. Navarro, a television personality, was assaulted in his home in January of 2014, apparently in retaliation for attempted rape. The incident stirred concerns over rape culture, as many questioned the motives of the female accuser. Twitter traffic reflected these issues. However, the American movie star Jennifer Lawrence also factored highly in the most searched items in the Philippines regarding the hacking of her private phone account.

Between March and July a vehemently anti-LGBT figure named Myles Munroe ranked among the most-searched topics in Nigeria[11]. The largest Nigerian event from the United States perspective (the kidnap of the Chibok schoolgirls in mid-April) factored only ninth in Nigeria. "Kidnapping" searches initially increased approximately four-fold in the month and a half following this incident, but then dropped to previous levels. Given the responsible group's influence and tactics, we suspect citizens may have been reticent to discuss any wrongdoing openly via social media.

South Africa's Google searches reflected considerable violence[12]. Among the top events driving traffic was the murder of a prominent soccer player in October during a burglary at his home. Oscar Pistorius also influenced the search patterns. The trial for the murder of his girlfriend began on March 3 and temporarily adjourned on May 20. The trial resumed on June 30 and lasted through August 8, with a judgment on September 12 and sentencing on October 21. At least two of the peaks in GBV-related Twitter traffic are coincident with these events.

These observations suggest that any quantitative analysis of social media traffic must control for current events, in order to separate fundamental trends from local variability. Furthermore, the patterns suggest that we will require separate models of GBV in social media, by country. Such control is well within the capability of computational analysis using topic-modeling approaches.

We note apparent correlations between certain countries: India and the U.S., as well as the Philippines and Nigeria at least for the bulk of the data. We observed a 48.1% overlap in the popular topics of the U.S. within the set for India for the month of August. Topics to determine overlap were based on the top 500 key-phrases (set of words in a message text) extracted using *tf-idf* based method (Nagarajan et al., 2009). Such content overlaps suggest the presence of an underlying, latent variable responsible for the observed relationship. Demographic properties of the countries investigated suggest the identity of this variable. We note a large Indian diaspora living in the U.S., potentially reacting to events originating in India. For example, the story of a

---

[10] http://www.google.com/trends/topcharts#geo=PH&date=2014

[11] http://www.google.com/trends/topcharts#geo=NG&date=2014

[12] http://www.google.com/trends/topcharts#geo=ZA&date=2014

15 year old Indian female acid attack victim received considerable attention following an Al Jazeera report in January (Dixit, 2014). The shared August peak coincides with an Independence Day speech by the Prime Minister of India (Narendra Modi) in which he urged suspension of the practice of questioning the families of girls regarding their social habits but never boys. Tweets with apparent origins in the U.S. may actually constitute an amplification of tweets originating in India, as the following tweet illustrates. Note, however, that the hashtags at the end are different, demonstrating the role of user engagement and amplification regarding conventional media.

India: *What world calls shining #India is the worst place for women in terms of #Rape http://t.co/F2oyKps60R #BanBollywood #MediaMafia*

US: *What world calls shining #India is the worst place for women in terms of #Rape http://t.co/zZoB17p1L1 #BanBollywood #PakMediaHijacked*

The U.S. and Indian patterns illustrate an important challenge in the interpretation of social media data—all trends do not necessarily reflect local events. Computational tools for monitoring GBV will need to conduct location analysis from the text to distinguish commentary about other countries. Model building can accommodate this distinction. However, more fundamentally, given the tweet pedigree, it is not clear to which country we should attribute the attitude.

*Table 4.* Percentages of sexual violence messages discovered by time slice and by country.

| Country | Phase-1 | Phase-2 | Phase-3 | ALL |
|---|---|---|---|---|
| United States | 65.10% | 64.78% | 63.90% | 64.44% |
| India | 66.05% | 65.41% | 65.30% | 65.45% |
| Philippines | 77.68% | 69.08% | 69.09% | 72.00% |
| Nigeria | 55.80% | 47.13% | 49.08% | 49.59% |
| South Africa | 71.15% | 66.96% | 63.69% | 66.24% |
| Total | 65.67% | 62.52% | 61.11% | 62.43% |

**Theme analysis.** Our original data collection categorized the tweets into three thematic groups: physical violence, sexual violence, and harmful practices. We used a rule-based classification approach where a tweet was considered relevant to a specific content theme (e.g., physical violence) if it contained any sequentially ordered pattern (e.g., woman.*dragged) based on the domain-expert provided keyphrases for that theme (as shown in Table 2). Sexual violence dominates the sample of GBV tweets (62% overall in the global corpus, in comparison to 5% for physical violence and 3% for harmful practices, where we found only a 12% overlap between sexual and physical corpuses, demonstrating a largely orthogonal categorization scheme.). We provide the distribution of sexual violence sample over time for the five countries in our study in Table 4. We note a general decrease over time with some variability between countries. As suggested above, we suspect early spikes related to incidents like the Vhong Navarro rape accusation in the Philippines and the initial publicity surrounding the Boko Haram atrocities in Nigeria.

**Sharing behavior analysis.** Social media provides the opportunity to distribute information, potentially reflecting both the senders' judgment of information importance, and reliance on the voice of others. Sharing functions to amplify these voices, often the voices of influential celebrities. We analyze two types of sharing behavior in the social media community surrounding GBV events: direct content resharing as a *retweet (RT)*, and indirect sharing via references to external resources, such as news, blogs, articles, and multimedia, using a *URL*, etc.

*Table 5.* Proportional RT and URL tweets by country.

| Country | RT | URL |
|---|---|---|
| United States | 45.47% | 42.25% |
| India | 46.68% | 42.63% |
| Philippines | 40.74% | 18.79% |
| Nigeria | 26.28% | 62.13% |
| South Africa | 44.25% | 31.49% |

We calculated the percentage of GBV retweets relative to the total count of tweets for each data sample as shown in Table 5. More than 40% of the GBV corpus is a retweet in the US, India, the Philippines, and South Africa, amplifying information that senders consider to be important. For comparison, Liu, Kliman-Silver, and Mislove (2014) found that retweets generally constitute just over 25% of the total volume of tweets. Although we note variability in retweet behavior between countries, the low retweeting frequency in Nigeria is particularly remarkable (see Table 5). One might hypothesize that a low literacy country such as Nigeria, in which senders are less able to compose messages, would have the highest retweet ratio. The adjacent analysis of the proportion of URL references with respect to the total corpus suggests a different sociocultural phenomenon at work concerning the identifiability of the responsible party. For GBV tweets containing URLs, Nigeria has the highest percentage of tweets with URLs in comparison to other countries. Numerous explanations can be tested, including literacy, credibility of the public press, and the possibility that reliance on external resources somehow reduces the threat of being identified as the responsible party.

**Author gender analysis.** We obtained gender identification for 37% of the users affiliated with the target countries (see Table 6). The reduced percentage is due to names missing in the Genderize API lexicon, as well as unconstrained natural language features of social media content, such as use of special characters in names, for instance, '@@shish' instead of 'Aashish', which is a male Indian name.

*Table 6.* Gender-wise distribution of data for the overall global corpus.

| Statistic | Users | % of Total Users | Generated Tweets | % of Total Tweets |
|---|---|---|---|---|
| **Total** | 3,036,576 | 100% | 13,942,592 | 100% |
| - Gender Filtered* | 1,148,329 | 37.82% | 2,771,686 | 19.88% |
| -- Female author | 563,016 | 18.54% | 1,324,292 | 9.50% |
| -- Male author | 585,313 | 19.28% | 1,447,394 | 10.38% |

*Filtered = where an author gender could be determined

Keeping in mind that statistically significant differences are certain with a large sample dataset, the distribution of gender appears approximately equal in Table 6. We also note a corpus tweet frequency average for a female author as 2.352 tweets per author, while 2.472 tweets per author for a male. However, based on the name classification procedure that we employed, Figure 3 separates the gender distribution for the examined countries, and creates an impression of gender inequality for the GBV content corpus. We note a discrepancy with Pew Research Center findings (2011) suggesting equal participation between genders in the United States. While GBV activists might be hiding their identities in developing countries such as Nigeria, this is an unlikely explanation for the observed US gender effect regarding the prevalence of GBV content.

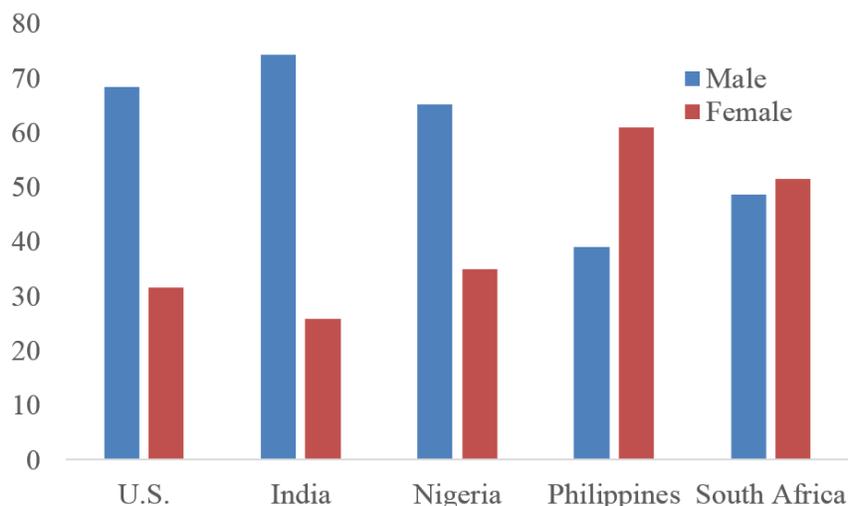

*Figure 3.* Relative distribution of tweets whose author's gender we could determine.

Literacy serves as a partial explanation for the observed ratios, except for the U.S. Apart from the explanation, opinions collected in the U.S., India, and Nigeria reflect a male bias, while opinions collected from the Philippines and South Africa are more balanced or even reflecting slight female bias. These observations have implications for the assessment of GBV attitudes, the general reach of anti-GBV campaigns using social media and the ability to target potential perpetrators and activists for engagement separately.

**Qualitative Content Analysis**

Thus far we have described the corpus with respect to dimensions that we anticipated at data collection: country, theme of GBV event, sharing methods, and gender. We have made suggestions about the surrounding context, including transient events that might explain the observed patterns. In the remaining analyses, we look more closely at message content for indications of GBV attitudes and behavior to clarify the requirements for future computational analysis capabilities.

**Language indicators.** Using Linguistic Inquiry Word Count (LIWC) software[13] (Tausczik & Pennebaker, 2010), we analyzed language of the content of all tweets generated by both genders. We used the predefined LIWC dictionaries that tally word frequencies in categories such as anger, sexuality, sadness, health, etc. Content corresponding to these particular categories, and in fact content across the majority of the standard LIWC categories appeared more frequently in tweets of male origin relative to tweets of female origin (significance at the level of $z > |1.96|$). However, we did note some LIWC categories in which females trended higher (*we*, *assent*, *impersonal pronoun*, *present*, *adverb*, *cause*, and *I*). As these are the unusual instances, they are the ones on which we will focus, leaving the potential category by country by gender interactions for future study. Consistent with the research on gender issues in communication (Tannen, 1996), female authors here are more collective and socially oriented. Their tweets call for action and are more likely to express or solicit agreement.

Female (South Africa): *@USER7 Absolutely. If we follow each other I can DM you my email address. I applaud your speaking out on the rape epidemic in SA*.

Female (Nigeria): *I am worried abt our approach to d fight against rape. Permit me to vent here. @USER8 @USER9 @USER10 @USER11 #CurbingRape*

Female authors are also more likely to provide opinion on causality, as in the following example tweet from India:
Female (India): *The major factor behind #Rape in #India is the #Bollywood which incites feeling 2 cross the moral limits #BanBollywood #GreaterPakistan*

Gender-specific analysis can be leveraged to design and promote anti-GBV campaigns. For example, tweets of female origin in India, although not guaranteed to be benevolent, could be amplified to extend their reach.

Following a manual review of a random sample of the corpus, we used computational analyses to describe the prevalence of attitude indicators across the whole corpus. Two of the attitude indicators examined here are the presence of humor indicators and GBV metaphors in sports. We also provide specific examples of tweet content from manual analysis of a random

---

[13] Information available here: http://www.liwc.net/

sample.

**Humor indicators.** Humor and related sarcasm might indicate a trivialization of an issue or an expression of underlying helplessness (Romero, 2010; Ancheta, 2000). We assessed the prevalence of humor references with related permutations of "haha" and "hehe". The number of such humor-flagged tweets by country appears in Table 7. The Philippines sample provides a greater proportion of such humor indicators by far. This is consistent with Filipino culture[14] in general. "Haha" and "ha" may play a different role than "He he", where the former constructions appear in conjunction with question marks and suggest sarcasm, while the later may not have these properties and function more like true laughter. For example, the following tweet from the Philippines questions the veracity of the plaintiff, pointing to an unrealistic timeline:

(Female): *RT @USER12: Panong mararape si roxanne cabaÒero kung nasa concert si vhong with vice? Haha! Tangina throwback rape na nga long distanÖ.*

The following tweet with its "Hehehe" notes that the slender appearance of the subject in an old photo creates the impression of a rape victim:

(Male): *Throwback!!! Nung ako'y naging isang payatot na rape victim. Hehehe http://t.co/va4qQStQcm*

Certainly, not all sarcasm will be marked with punctuation and laughter icons, as in the following example:

(Female, Philippines): *RT @USER13: Rape prevention nail polish sounds like a great idea but Iím not sure how youíre going to get men to wear it*

Presenting similar challenge to computational identification, the following example suggests mock pride in the Philippines as a record holder for the world's fastest rape, but without the punctuation and laughter indicators.

(Female, Philippines): *@User14 #PROUDPinoyKasi hawak pa rin ng Pilipinas ang Guinness Record para sa pinakamabilis na rape! Woohh! Thanks Deniece! NakakaprouÖ*

Such constructions are not as compatible with screening searches based on lexical items and punctuation. We interpret the lexical markers as potentially correlated indicators for jokes that would be much more difficult to identify, not only with respect to semantic processing but also cultural nuance. Furthermore, while real semantic differences may distinguish the usage of "ha" and "he" constructions, including nuances related to length, we need not make such distinctions in order to call attention to the prevalence of these constructions and their general implications for GBV attitudes and behavior, regarding the veracity of the plaintiff or the trivialization of rape. Nevertheless, due to cultural differences in the role and expression of humor, we cannot advocate direct comparison between countries. Our keyword approach may underestimate the prevalence of humor outside of the Philippines. Moreover, we suspect the need for different humor indicators and models between countries. However, we do suggest that the prevalence of GBV humor may provide a useful metric for changes in attitude over time within a region.

---

[14] A brief overview of traditional Filipino values can be found here:
http://en.wikipilipinas.org/index.php?title=Philippine_Core_Values

The observed female bias in the Filipino sample reinforces the interpretation of humor as an expression of helplessness.

(Female): *@User15 no place is /THAT/ safe and I don't really want to take the risk especially when I'm alone :((( rape and robbery are very"*

In support of this interpretation, we note the prevalence of the following construction: "baka ma rape", meaning "you might get raped".

(Male): *#XXXXXX mag ingat ka huh? Wag ka masyado mag pagabi! Baka ma rape! Sige ka! -.-*

We searched on this construction in the Philippines data as well as its English counterpart "you might get raped" in the other countries with the following results. We identified 717 tweets originating in the Philippines warning of the rape potential. Similar language exclusively in English uncovered tweet frequencies ranging from a high of 74 in the US to a low of 46 in South Africa. The Philippines, with less than one third of the population of the US generated nearly ten times more warnings of potential rape. This finding reinforces the need for a context and culture-specific models to study GBV behavior.

*Table 7.* Analysis of humor-flagged tweets indicated by permutations of humor indicators.

| Country | Flagged tweets | Total tweets | Percentage |
|---|---|---|---|
| Philippines | 10418 | 105550 | 9.870% |
| South Africa | 1801 | 104430 | 1.725% |
| USA | 8067 | 698077 | 1.156% |
| Nigeria | 1419 | 134671 | 1.054% |
| India | 5195 | 549033 | 0.946% |

*Note.* Lexicon set defined as *{haha, ha ha, hehe, he he, he he he, hahaha, lol, lmao}*.

**GBV and sports metaphors.** Sports involve competition, and violent metaphor is a common device in sport discussion (Lewandowski, 2010, 2011, 2012). GBV metaphors may appear in a sports corpus as an indication of dominance, such as the following example from the South African tweet corpus:

South Africa: *The German Team is on a Steroid-induced anal-rape rage against Brazil right now.*

Comparing the prevalence of GBV metaphors in sports across countries is difficult because a given sport does not capture national attention equally across countries. We obtain some insight regarding the GBV metaphor in tweets related to sports from exchanges concerning the FIFA World Cup contest, held from 12 June through 13 July 2014. Soccer is the most popular sport in Nigeria and South Africa, one of the top three favorite sports in India, and a top ten favorite in the Philippines and the U.S.[15]. While all of the countries we examined have eligible teams, only the U.S. (ranked 13) and Nigeria (ranked 44) participated in the final tournament[16].

We should not expect any FIFA-related content in a corpus designed to capture GBV issues. Yet, Table 8 illustrates the existence of tweets flagged for containing references to: *football, futbol, soccer, worldcup, world cup, fifa, fifacup,* and *soccercup,* as well as pairs of team names participating in the tournament. Although the percentages are small, every country examined provided such instances in a ranking consistent with soccer popularity, with the exception of the U.S. The finding requires a nuanced interpretation. On the one hand, rape as a metaphor suggests a trivialization of the primary definition. On the other hand, at least one dictionary[17] indicates an archaic definition meaning plunder or violation, for example regarding the environment. On the one hand violent metaphor is a common device in sport discussion. On the other hand, it is just one of many common metaphors (Lewandowski, 2012). In fact, in Lewandowski's (2010-2011) extensive analysis of violent, conflict-oriented metaphor in football and in sports journalism, none of the 551 violent metaphors invoked rape.

A difference in the popularity of sports between countries is only one challenge to direct comparison of GBV metaphor between countries. The development of a knowledge base to support analysis poses further challenge since different regions follow different teams with different players. As in the interpretation of humor, we believe the best use of sports metaphor data is to provide indicators of attitude trends over time by region, while adjusting for seasonal variations in sporting events or averaged annually.

*Table 8.* Analysis of sports-related tweets indicated by sports lexicon.

| Country | Flagged tweets | Total tweets | Percentage |
|---|---|---|---|
| United States | 1053 | 698077 | 0.0015% |
| India | 657 | 549033 | 0.0011% |
| Nigeria | 320 | 134671 | 0.0023% |
| Philippines | 47 | 105550 | 0.0004% |
| South Africa | 214 | 104430 | 0.0020% |

*Note:* FIFA World Cup sample for each country in the period of June 12 to July 13 2014.
Lexicon: { *football, futbol, soccer, worldcup, world cup, fifa, fifacup, soccercup, [country-names pairs from http://en.wikipedia.org/wiki/2014_FIFA_World_Cup]* }

**Policy-making and intervention insights via manual analysis.** Our goal is to generate attitude metrics and inform mitigation campaigns automatically. Indeed, computational analysis informed all of the above examples. In contrast, for the following examples, we manually examined random subsets drawn from each of the three slices of our data corpus, as 200, 500,

---
[15] http://mostpopularsports.net/football-soccer-popularity
[16] http://en.wikipedia.org/wiki/2014_FIFA_World_Cup
[17] http://dictionary.reference.com/browse/rape

and 500 tweets. We end our presentation of GBV data by indicating the kind of content available in the corpus compatible with somewhat more specialized computational analysis. Developing the necessary content-specific filtering methods (feasible using domain knowledge of politics, business, and behavioral theories) is particularly critical for appropriately routing content to specific policy making recommendation agencies. Some examples follow below:

a.) Behavior pertaining to government/public officials/leaders
- *@USER16 NCP leader doesn't know rape happens due to pervert mindset . what abt unreported rapes happen with minors & toddlers.*

b.) Commercial considerations
- *RT @USER17: .@USER18 Please don't allow violent hatemongers use your app to harass and exploit marginalized women. \"http://t.co/DoRkbXFuN…*
- *Spring Breakers isn?t just a terrible movie, it reinforces rape culture http://t.co/pxTAZftM8v*
- *RJ Police already said that it's not a Rape case!But Media neglected it 4 creating SPICY news!#KnowTheTruth & Rise4Justice*

c.) Persuasive/encouraging message content
- *RT @USER19: Pregnancy, periods, breast cancer, being walked on, rape, harassment, abuse; females go through a lot. WOMEN ARE STRONG.*
- *@USER20 @USER21 Is it acceptable to use gang rape in an example?*

d.) Stereotype association
- *RT @USER22: It is not my job to coddle and "educate" young Black men when it comes to violence against women. Y'all wanna "teach"?*

Automated filtering of GBV content according to such dimensions is challenging but feasible given the appropriate knowledge bases.

**[4.] DISCUSSION AND FUTURE WORK**

Here we revisit the limitations of conventional, survey-based methods for monitoring GBV attitudes to discuss the progress we have made, the limitations we face, and the remarkable feasibility of further advances in Computational Social Science to address the persisting limitations.

**Progress**

We noted above the limitations of conventional methods in gathering GBV data for specific regions. Data are either highly aggregated at the country level or missing. We have provided substantial amounts of GBV data for all five of our target regions. User profile location and tweet origin location metadata provide the capability of identifying content specific to much smaller units of analysis, e.g., type of socio-economic region or even city (Leetaru et al., 2014). This capability supports targeted GBV campaigns, where the prevalence and types of violations may vary. Below we note progress with respect to a number of additional concerns regarding conventional data gathering methods.

**Reducing sampling bias.** We noted the presence of bias in survey methods due to artificial sample selection. We have overcome some of the bias concern simply by the scope of data collection that is feasible with social media. We have the opportunity to observe male and female attitudes by specific regions, over time. However, bias does remain in our data collection methods. Some of the persisting bias is amenable to adjustment. For example, bias exists in the form of literacy assumptions by country and by gender within country and Internet penetrability. Awareness of these sampling issues allows us to amplify the weight of content from underrepresented participants, e.g., the contributions from females in lower penetration areas. This is possible because we can assess both gender and location in the available data. Of course, we cannot amplify content that does not exist. This limitation is fundamental, but it also identifies the regions where this limitation occurs and where we should focus other methods of data collection. We do acknowledge the existence of automated bot user accounts on Twitter (Zhang & Paxson, 2011) that could have affected our data collection. Although Twitter has been actively working on controlling spam accounts, and given the sensitivity of GBV topic, we suspect lesser contribution from such bot accounts in the GBV related content. Future work is required to filter bot-generated content in our datasets.

**Reducing content bias.** We suggested that survey participation itself constitutes a form of bias. Apart from the sampling issues noted above, the posted content on social media itself reflects a form of bias. Participants are still providing what they consider to be socially acceptable content. This is particularly apparent in the absence of commentary regarding Boko Haram and GBV in Nigeria. However, this observation alone can inform the course of action for policy makers, e.g., publicizing sensitive topics. Nevertheless, the available social media are not responses to external (survey) queries about attitudes. Instead, these express attitudes directly, making them susceptible to analysis.

**Analysis speed.** We have over 1.5 million social media postings collected in 2014 from just five countries to support the claim that GBV issues generate commentary. Moreover, we have completely eliminated the need for labor-intensive data collection, and as a result have overcome the cost and lag limitations in data collection. We produced analysis within a year in contrast to survey methods, despite the absence of complete automated capability. This rapid turnaround enables the development of dynamic metrics to assess the results of campaigns designed to curb GBV. A rapid measurement capability can play a role in the promotion of effective efforts and the abandonment of those that are less effective, with real consequence to the alleviation of human suffering.

**GBV attitude and behavior metrics.** Survey items tend to address attitudes but not behavior. Social media data provide both attitudes and behavior, inasmuch as jokes and metaphor are both behavior and attitude. This provides us with potential measures of tolerance for GBV. Thus, the editing of socially acceptable content that constitutes a form of bias in data collection is the very same behavior that tells us what is considered acceptable. This provides a means to measure the effectiveness of anti-GBV campaigns, both those directly targeted by the potentially offensive jokes and metaphor as well as those targeted by specific but apparently unrelated concerns such as the effect of holidays or weather.

**Survey methods are, by design, static.** Standard measures purport to provide evidence that is comparable across time and regions. However, standardization ignores the effect of context. Social media trends over time tell us that context cannot be ignored. For example the publicity surrounding a celebrity involvement in a GBV-related event spikes social media commentary. The availability of such events (Tversky & Kahneman, 1973) may very well influence survey responses. But standalone surveys have no way to account for this influence. Our computational social science methods allow us to complement the interpretation of GBV commentary with adjustments for the influence of events at a short-time scale in order to discern long-term trends.

Finally, and completely outside the typical survey content, the analysis of social media also promises to assist GBV campaigns, both by targeting the views of specific groups and by providing content recommendations regarding law enforcement, politics, health services, and commerce.

**Limitations & Challenges**

All data collection methods suffer from limitations, and ours is no exception. Here we note several concerns, indicating which are amenable to computational solution and which are more fundamental to all data collection and interpretation methods.

**Unconstrained natural language text.** Our keyword-based crawling limits the completeness of the resulting corpus. Given the natural language of social media messages, we cannot guarantee collection of every single relevant message. Moreover, keyword selection matters. Countries vary the terminology they employ for different purposes. For example the word "rape" in tweets that originate in India generally refer to events in India, but "sexual assault" in the Indian corpus returns mostly American incidents. We are further constrained at present by restriction to the English language. The common word 'rape' in Tagalog and English, along with laughter indicators enabled analysis of the Philippines database. Furthermore, our dependence on keywords glosses over the different definitions of rape across cultures.

**Global event sensitivity.** Both a feature and a limitation, we demonstrated the influence of events on social media content, both within a region and between regions. World events provoke the articulation of public opinion and provide an unprecedented large-scale opportunity to gather opinion and attitude. At the same time world events create variations in magnitude that require adjustments to frequency counts, in order to reflect enduring trends over time.

**Inter-regional comparison.** We commented earlier that differences in the legal definition of GBV hindered comparison of GBV rates between countries. While we believe our measurements within a region can be informative about attitude change over time, the ability to compare GBV issues between regions still poses substantial challenge. Our two proposed measures, sports metaphors and jokes illustrate the challenge. A given sport is not equally important across countries (or even regions), so that the prevalence of more frequent violent sport metaphor in one country relative to another may simply reflect population interest in the sport. Allowing type of sport to vary between regions confounds sport with region, so that we might be learning more about issues with the sport than the region.

The meaning of jokes notwithstanding (sarcasm, trivialization, or helplessness), we cannot compare the prevalence of GBV jokes between countries without factoring in the prevalence of jokes between countries in general. Thus, we have not escaped the local, sociocultural influences on measures that prevent between region comparisons. Our contribution to this issue is to establish that it is not a limitation of specific measures such as police reports and surveys. All measures reflect these sociocultural influences.

**Location identification.** Location references of the messages are not always consistent with the location of the author profile or GPS coordinates of the source device. U.S. events appear in the opinions of people in other countries, and vice versa. The technical aspect of this problem resolves with techniques that discern event location from the message text or ancillary URL content. This is not necessarily without remaining uncertainty, as people often assume shared context location identification, e.g., with abbreviated names for familiar landmarks resulting in referential ambiguity. However a far greater concern lies in the attribution of attitude when the content corresponds to a remote event.

**Correlational logic.** Our argument has a correlational component. For example, we can identify indicators of humor such as "hehehe" and "hahaha" using computational methods. However, automatically detecting every instance of a joke embedded in social media content is not computationally feasible. We assume that the frequency of humor keywords in a GBV context correlates with the frequency of GBV jokes without such keywords. Although this assumption is worth confirming using a manual classification of jokes, we note that the presence of laughter indicators alone in this context is potentially offensive.

**Future Work**

We demonstrated the potential of social media to inform policy makers regarding attitudes and behavior, measuring the effect of campaigns, and even providing campaign concepts. However, much technical, theoretical, and practical work remains to realize the full potential of the medium for the GBV application.

Technical advances are required in order to distinguish message origins from message content and support more comprehensive gender detection. Substantial work remains in the development of specific knowledge bases to guide the detection and interpretation of jokes as well as metaphor and commentary directed to particular entities such as government and business. The knowledge bases must be dynamic to capture the transient events that mask fundamental attitude. Theoretical guidance from social science is required to attribute humor and metaphor to either despair or tolerance or the lack thereof. The interpretation of retweets across regions raises the problem of attribution in more than a practical sense. The issue is whether and how to weigh the endorsement: as a property of the original sender, or the endorser who is amplifying the message. Does the endorsement reflect opinion of the endorser's location independent of birth origins? Comparison of GBV threats between countries, though raising measurement issues, is not an exclusively methodological problem. Instead, it is a sociocultural and political issue, concerning the articulation of globally established norms to determine the deployment of global resources. We lack the policy expertise to weigh in on such matters.

However, we do endorse the development of adaptive regional, and even local models for GBV behaviors and attitudes. Although a large model encompassing all countries holds a certain appeal, it is fallacious to assume that all regions have the same underlying issues and beliefs. Local socio-economic conditions, literacy, Internet penetrability, gender, crime rates, religious beliefs, liberty, and many other unexamined but constant factors influenced the data and their interpretation.

**Data accessibility for policy-makers.** Near-real-time information benefits an urgent need to reduce GBV and the associated suffering. Conventional methods of disseminating public opinion are reflected in written reports covering multiple years, and issued with considerable delay following data collection. However, governments, NGOs, International Organizations, Aid Workers, etc. require far higher bandwidth access to the dynamic data and analysis in order to measure current public opinion and the effect of anti-GBV campaigns.

The Twitris collective social intelligence platform (Sheth et al. 2014) provides a foundation for delivering the necessary high bandwidth access. Twitris supports the presentation of thematic data along spatial and temporal dimensions (Nagarajan et al. 2009), network relationships among people related to the distribution of content (Purohit et al. 2012), and sentiment (Chen et al. 2012) as well as real-time trends[18] shown in Figure 4.

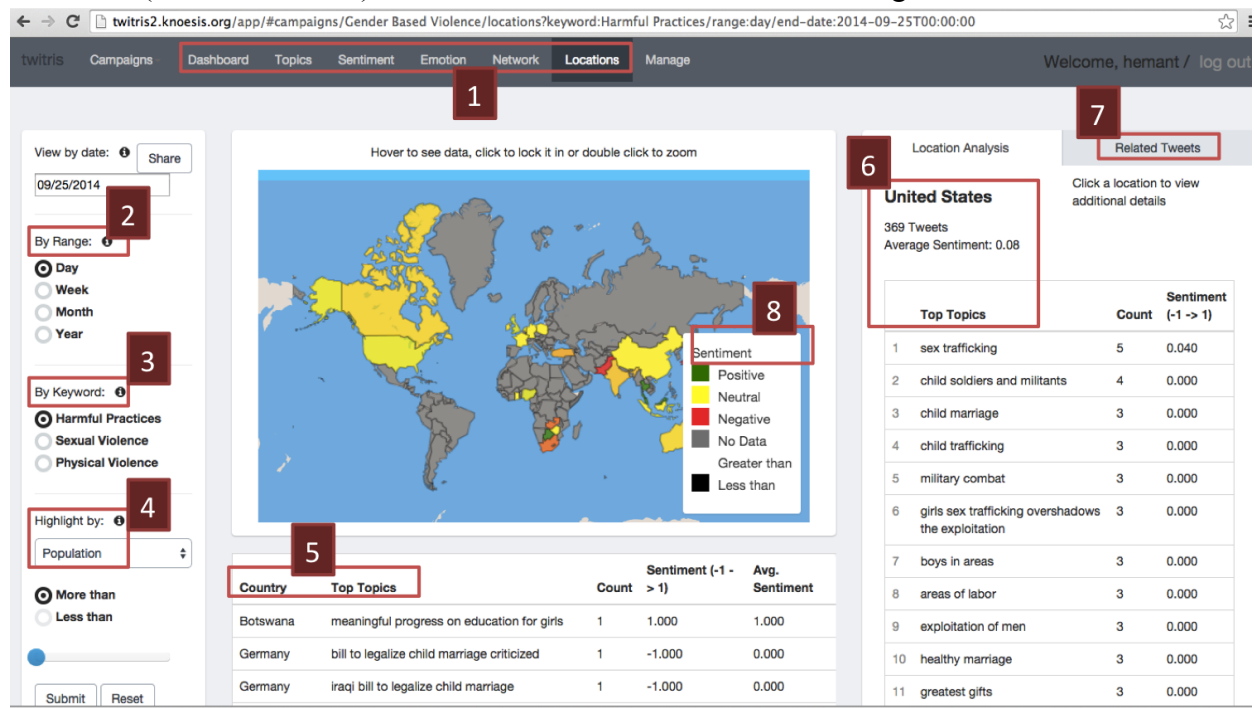

*Figure 4*. A snapshot of the Twitris social analytics platform showing ongoing research to monitor and analyze the GBV data by a variety of dimensions: location (1, 8), time (2), types of content theme, such as sexual violence (3), demographics-based map visualization updates (4), top topics by country or region of interest (5), country-specific demographics including top topics and related tweets (6, 7), sentiment-based heat map (8), as well as fine-grained analysis of emotion expressed, who-talks-to-whom network for

---

[18] A currently monitored GBV campaign is available at:
http://twitris2.knoesis.org/app/#campaigns/Gender%20Based%20Violence/dashboard Register for full dashboard view.

influential users in GBV topics, and a real-time trends dashboard (1).

Twitris analyzes a topic such as GBV, and provides real-time, scalable analyses of social data streams for greater insights and actionable information to improve intervention. For example, Twitris can monitor the real-time public sentiment and emotional reaction to criminal reports and justice system response by region, permitting side-by-side comparison with other events. Under user guidance, Twitris automatically identifies the key topics meriting further attention. Twitris' network analysis assists in the measurement of anti-GBV campaign diffusion to measure campaign effectiveness, as well as the identification of users who will spread targeted campaigns.

**CONCLUSION**

The study presents the case for the advancements in Computational Social Science to inform GBV policy design and anti-GBV campaigns. Big social data complement more controlled but slower survey-based data collection and analysis methods, whose conclusions may become obsolete in a dynamic world that continuously generates noisy data responding to transient events. The lag in surveys and the noisy nature of data limits the use of conventional methods for measuring the effects of changing GBV attitudes and anti-GBV campaigns. Computational Social Science supports the collection and analysis of a large GBV corpus. To demonstrate the promise of Computational Social Science for this social issue, we analyzed nearly fourteen million Twitter messages collected over a ten-month period. We demonstrated an ability to exploit a large sample to reduce bias relative to conventional data collection. We demonstrated an ability to examine data by region and gender, identifying content such as humor and metaphor that have implications for both the measurement of GBV attitudes as well as specific targets for anti-GBV campaigns. Our methods constitute an inexpensive way to engage with citizens at unprecedented scale, including the collection of public views regarding the behavior of government and business to revolutionize the conduct and measurement of anti-GBV campaigns.

**ACKNOWLEDGEMENT**

We are thankful to our colleagues at the United Nations Population Fund NYC—especially Upala Devi, Judy Ilag, and Maria Dolores Martin Villalba—and the Kno.e.sis Center, especially Lu Chen, current research interns Kushal Shah and Garvit Bansal from LNMIIT India, and Maria Santiago for invaluable continued support in discussion and review of our GBV research.